\documentclass[aps,prb,twocolumn,showpacs,preprintnumbers,floatfix,superscriptaddress]{revtex4}

\usepackage{graphicx}
\usepackage{dcolumn}
\usepackage{subfigure}
\usepackage{wrapfig}
\usepackage{cancel}
\usepackage{color}
\usepackage{textcomp}
\usepackage{amsmath}
\usepackage{amsfonts}
\usepackage{amssymb}
\usepackage{array}

\newcommand{\ie}{\it{i.~e.}}

\newcommand{\etal}{\it{et~al.}}
\newcommand{\Tc}{$T_{\text{c}}$}

\begin{document}

\title{Distinct electride-like nature of infinite-layer nickelates and the resulting theoretical challenges to calculate their electronic structure}
\author{Kateryna Foyevtsova}
\author{Ilya Elfimov}
\author{George A. Sawatzky}
\affiliation{Department of Physics \& Astronomy, University of British Columbia,
Vancouver, British Columbia V6T 1Z1, Canada}
\affiliation{Stewart Blusson Quantum Matter Institute, University of British Columbia, Vancouver, British Columbia V6T 1Z4, Canada}

\date{\today }

\begin{abstract}
We demonstrate in this paper that the
recently discovered infinite-layer (IL) nickelates
have much in common with a class of materials known as electrides.
Oxide based electrides are compounds in which topotactic removal of
loosely bound oxygens leaves behind voids with a landscape of attractive
potentials for electrons.
We show that this is also what happens in
the IL nickelates, where one of the
two electrons (per formula unit) freed during the topotactic synthesis
is to a large
degree located in the oxygen vacancy position,
occupying partially a local $s$-symmetry interstitial orbital, rather than
taking part alongside the other electon in converting Ni from 3+ to a full 1+ oxidation state.
On the other hand,
the involvement of the rare-earth $5d$ states is found to be rather
indirect.
Our study offers a new and detailed perspective on
the mechanisms through which the presence of the interstitial charge
can determine the superconducting and other properties of the IL nickelates.
To this end, we demonstrate that the interstitial orbital in question,
referred to by us as the zeronium~$s$ or Z~$s$ orbital,
forms strong covalent bonds with neighboring Ni~$3d_{3z^2-r^2}$
orbitals, which in turn facilitates the one-dimensional-like dispersion
of the Ni~$3d_{3z^2-r^2}$ band along the $c$-axis direction,
leading also to a possible large out-of-plane coupling between Ni magnetic
moments.
This finding, reinforced by our electron localization
function analysis, points to a fundamental
distinction between the nickelates and the structurally analogous
cuprates, may explain the absence of superconductivity
in hydrogen-poor samples,
and is certainly in agreement with
the observed large $z$-polarized
component in the Ni $L_3$-edge x-ray absorption spectra.
In addition, by using DFT+U calculations as an illustration,
we show that the electride-like nature
of the IL nickelates
is one of the main
reasons
for the theoretical difficulty in determining the much debated elusive
Fermi surface of these novel superconductors
and aslo in exploring the possibility of them
becoming excitonic insulators
at low temperatures.

\end{abstract}

\maketitle
\section{Introduction}
The study of the recently discovered Ni-based
high transition temperature (\Tc) superconductors
with an infinite-layer (IL) crystal structure \cite{Li19,Osada20,Zeng22,Pan22}
promises to provide important clues to our understanding of unconventional
superconductivity in complex oxides.
Such optimistic expectations
are primarily fueled by
the intriguing
similarity that
the nickelates bear with the famous superconducting cuprates \cite{Bednorz86,Keimer15}.
In both, the two-dimensional NiO$_2$ or CuO$_2$ layers constitute an essential part
of the crystal structure [see Fig.~\ref{fig_nmdft}~(a)],
the transition metal (TM) ion is found in, formally,
the same $d^9$ configuration, and a superconducting
dome develops upon hole doping within similar doping ranges.
As, however, intense research efforts rapidly expand
our knowledge of the nickelate superconductors,
important differences are becoming apparent,
too (see reviews \cite{Chow22,Nomura22,Gu22,Zhou22_r,Botana21}
and references therein). Thus, unlike the cuprates,
the parent compounds of the IL nickelates behave
as bad metals and, despite the evidenced
strong short-range magnetic correlations \cite{Fowlie22,Ortiz22,Zhou22,Lu21}, seem to have a charge \cite{Rossi21} rather than
spin \cite{Ortiz22} ordered low-temperature state.
Also, as the most recent experimental
study reveals, superconductivity occurs only in samples
with substantial hydrogen concentrations within a narrow range
of 22 to 28 percent\cite{Ding23}.

These observations are in line with the conclusion
from an early theoretical study of the nickelates' electronic structure
\cite{Lee04} that,
notwithstanding their identical formal $d$ electron count,
``Ni$^{+}$ is not Cu$^{2+}$'',
implying also a possibly different superconducting pairing mechanism.
One of the major distinctions originates from the fact that the charge-transfer energy
between the TM $3d$ and oxygen $2p$ states is
considerably larger in the nickelates \cite{Jiang20},
which, seemingly, justifies placing these systems
in the Mott-Hubbard rather than charge-transfer
regime of the Zaanen-Sawatzky-Allen (ZSA) classification scheme \cite{Zaanen85}.
Yet, a simple Ni~$3d_{x^2-y^2}$ orbital
based Mott-Hubbard picture is inconsistent with
many of the key experimental features of the parent IL nickelates either.
One obvious such feature is their metallic behavior
that is likely to be related with the presence of two small electron pockets
of mixed orbital characters,
one at the Brillouin zone center and one at
its corner, which are routinely seen in electronic structure
calculations \cite{Nomura22,Been21}. It should be noted, though, that
a full and precise determination of the nickelates' Fermi
surface remains challenging to both theory and experiment.
Also at odds with the single orbital scenario is the fact
that the x-ray absorption spectra (XAS) at the Ni $L_3$-edge
have a strong $z$-polarized component, clearly indicating
a presence of Ni~$3d_{3z^2-r^2}$ holes in the ground state of the parent compounds \cite{Rossi21_xas}.
Such observed partial occupation
of the $3z^2-r^2$-symmetric Ni~$3d$ orbital in addition to the
$x^2-y^2$ one
makes is rather challenging to describe the IL nickelates using
the ZSA classification scheme,
since, in order to keep the unit cell charge neutral,
there must be electrons in formerly unoccupied states of local symmetry that can mix with the $3z^2-r^2$ symmetry at the Ni site, which, however,
are not provided by the atomic orbitals of the $R$NiO$_2$ crystal structure.

Given therefore that the IL nickelates
seem to transcend the standard ZSA paradigm and also motivated
by the recently suggested critical and intimately intertwined roles of the hydrogen and
the interstitial charge for the nickelates' superconductivity,
in this paper we take an alternative look at their electronic structure
by considering
the electride concept frequently encountered in
oxides prepared by a topotactic removal of selected oxygen ions \cite{Zhang15,Matsuishi03}.
Scientific interest in electride materials has recently
been on a rise \cite{Liu20} and is driven by their exotic nature
and a range of interesting physical properties they display,
including superconductivity \cite{Zhang17}. Electrides consist of
structural voids coordinated by mainly positive cations where
unbound electrons resulting from the oxygen removal can get
trapped by a strongly attractive potential.
Rather than occupying the cation bands, these electrons are relatively
localized in the voids and act like anions of charge $-1$ and spin $1/2$.
An analogy with the IL nickelates can readily be drawn by considering
the topotactic synthesis of the latter from the precursor rare-earth nickel
perovskites $R$NiO$_3$ ($R=$ La, Nd, or Pr), in which Ni ions assume a formal
valence of $3+$.
The extracted apical oxygens leave behind two electrons
per formula unit which should, according to conventional thinking,
convert Ni$^{3+}$ into Ni$^{1+}$. However, it is quite feasible that
whereas one of the two doped electrons is indeed on Ni,
forming a very stable $d^8$ state with spin 1 as in NiO,
the second electron is localized in the void at the O
vacancy site which indeed is surrounded by four Nd$^{3+}$ and two Ni ionic neighbors,
with the actual ionic charges of the latter being closer to 2+ than to 1+
in our proposed electride scenario.

In this paper, we use density functional theory (DFT)
calculations to study the possibility of this actually occurring in NdNiO$_2$,
a representative parent IL nickelate,
by explicitly introducing a fictitious atom without a nucleus at the oxygen
vacancy site, which is represented by a large white sphere in Fig.~\ref{fig_nmdft}~(a),
and centering a set of basis functions around it.
Without changing any physics,
this allows us to analyze the electronic structure of NdNiO$_2$
in terms of local $s$, $p$, and $d$ symmetries.
We will call this fictitious atom ``zeronium'' becasue of its  effective zero nuclear charge,
a term coined in the late 1970s \cite{Rompa84}, and use a chemical symbol “Z” for it.
We note that a variety of alternative terminologies exist.
Thus, the electron occupying (either fully or partially) an interstitial orbital inside the void
is also referred to in the literature as an interstitial, anionic,
or electride electron \cite{Liu20,Dale18} or as an interstitial quasiatom \cite{Miao15}.
As will be shown,
the charge density of the Ni~$d_{3z^2-r^2}$ character extends well into the interstitial region in the $c$ direction, and by including
Z we can quantify this in the same way that we can quantify the covalent mixing of the Ni~$d_{x^2-y^2}$ orbital with the linear combination of the neighboring O~$2p$ orbitals of
the $x^2-y^2$ symmetry in the Zhang-Rice singlet\cite{Zhang88}.
It will be emphasized that the hybridization between the Ni~$d_{3z^2-r^2}$
and the Z~$s$ orbitals is by far more important
in the interpretation of the low-energy electronic structure
than the symmetry-restricted involvement of the Nd~$5d$ and O~$2p_{\pi}$/$2p_z$
states, which have been at the focus of most previous theoretical studies.

The structure of the paper is as follows.
In Section II, we will describe our computational methods and provide
details of our electronic structure calculations.
In Section III, we will present our computational results,
first demonstrating the importance of the hybridization
between the zeronium~$s$ orbital
and its two neighboring Ni~$3d_{3z^2-r^2}$ orbitals along the $c$ axis.
The relatively secondary roles of
the Nd~$5d$ and O~$2p_{\pi}$/$2p_z$ states will be
illustrated by means of a computational experiment
involving elimination of the Z~$s$ states from the vicinity of the Fermi level.
We will then use the electron localization function (ELF) to show
the real space localization of the electride ({\it i.e.}, zeronium~$s$)
electron and compare NdNiO$_2$ with the iso-strictural CaCuO$_2$, finding the two of them
to approach qualitatively very different electronic ground states in the limit
of an exact-exchange ({\ie}, Hartee-Fock) treatment.
Further, we will show that the popular DFT+U method is seriously
challenged in the presence of the strong covalent bonding
between the zeronium~$s$ orbital and the Ni~$3d_{3z^2-r^2}$ orbital,
by highlighting qualitative inconsistencies between the results of DFT+U
calculations obtained from different DFT codes.
It will be emphasized that, since there is no doubt that these materials are strongly correlated, in the end it will be model Hamiltonians with methods like
dynamical mean-field theory (DMFT) or exact diagonalization or impurity like approximations that are needed to describe the experimental data.
In Section IV, drawing support from our computational results
and from the recent impurity calculations by Jiang~{\etal}\cite{Jiang22}, we will
discuss a realistic electronic configuration for Ni in the parent IL nickelates,
in which the zeronium states play one of the key roles
giving rise to such strongly contributing configurations
as $d^{8,\, S=1}_{x^2-y^2,\, z^2}\, s$ and $d^9_{z^2}\underline{L}_{x^2-y^2}\, s$,
where $s$ denotes an electron in Z~$s$ and $\underline{L}$
a hole in O~$2p$. This will be followed by a short discussion of the possibility
for the nickelates to transition into an excitonic insulating state at low temperatures.
Conclusions will be offered in Section V.

\section{Methods}
We consider primitive unit cells of NdNiO$_2$,
with the lattice constants $a=3.91$~{\AA}
and $c=3.37$~\AA,
and of CaCuO$_2$, with the lattice constants $a=3.86$~{\AA}
and $c=3.20$~\AA.
When not specified otherwise, our DFT
calculations are performed using the augmented plane wave all-electron
package WIEN2k\cite{wien} and choosing
the gradient-corrected
local density approximation (GGA)\cite{Perdew96} for the energy functional to treat the exchange and correlation effects. We use $RK_{\text{max}}=7.0$ and a $12\times12\times14$ $\Gamma$-centered $k$-point grid
for bandstructure calculations but increase the $RK_{\text{max}}$ value to 9.0 for calculating ELF. The atomic muffin-tin radii $R$
are set as $R$(Nd)=2.5, $R$(Ni)=1.57, $R$(O)=2.12, and $R$(Z)=1.61 in bohr
[$R$(Nd)=1.32, $R$(Ni)=0.83, $R$(O)=1.12, and $R$(Z)=0.85 in \AA].
The three Nd~$4f$ electrons are treated as core electrons.
For the two-formula-unit supercell with the $G$-type ordering of Ni spins
considered in Section~III~C, we use a $10\times10\times10$ $\Gamma$-centered $k$-point grid
and keep the same values for $RK_{\text{max}}=7.0$ and for the muffin-tin radii.
We find that it is important to use the tetrahedron rather than smearing method
for integrating over the Brillouin zone\cite{Santos22}, since otherwise the Fermi level
can be determined incorrectly resulting in a loss of the hole Fermi pockets,
which would obviously violate the charge neutrality principle.

In addition to WIEN2k, two pseudo-potential codes, VASP\cite{Kresse96}
and Quantum Espresso\cite{Giannozzi09,Giannozzi17}, are also used
in order to calculate the ELF of NdNiO$_2$ and CaCuO$_2$
within the hybrid functional method and/or
the Fermi surface of NdNiO$_2$ within the DFT+U method.
In these calculations, the energy cut-offs are set to ENCUT=320~eV
and ecutwfc=60~Ry in VASP and Quantum Espresso, respectively.
In the case of hybrid functional calculations on the
primitive unit cells of NdNiO$_2$ and CaCuO$_2$,
the size of the $k$-point grid is reduced down to $6\times6\times6$.
%

The Wannier functions based analysis
of the bandstructure of NdNiO$_2$ is performed by using the maximally
localized Wannier functions (MLWF) method\cite{Marzari97} as implemented in
the WANNIER90 code\cite{Pizzi20}.

\section{Results}
\subsection{Zeronium~$s$ orbital without magnetism and correlations}

\begin{figure}
\begin{center}
\colorbox{white}{
\includegraphics[width=0.47\textwidth]{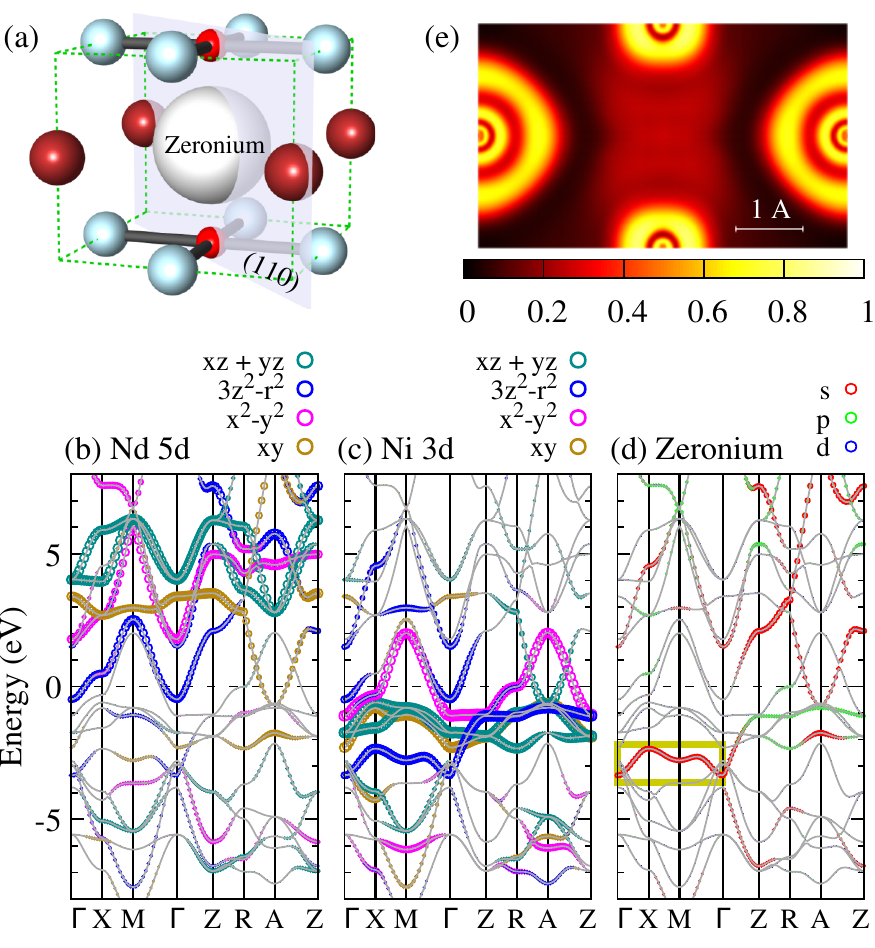}
}
\caption{(a) Crystal structure of infinite-layer nickelates
$R$NiO$_2$ ($R$=rare-earth atom).
Different atoms are represented by spheres as follows:
$R$ = large red, Ni = small red, O = blue, O vacancy (zeronium) = large white
at the center of the unit cell. (b) and (c) Band structure of NdNiO$_2$
from a non-spin-polarized GGA calculation projected onto
atomic (b) Nd~$5d$ and (c) Ni~$3d$ orbitals. Orbital character in a
given state is proportional to the area of the circle of
a respective color. (d) Same NdNiO$_2$ band structure as in
(c) and (b) but projected onto the zeronium orbitals
with $s$, $p$, and $d$ symmetries.
The yellow rectangle highlights
the bonding combination of the Ni~$3d_{3z^2-r^3}$ and Z~$s$ orbitals
at $k$-vectors with $k_z=0$. The zero of energy is at the Fermi energy.
(e) Electron localization function (ELF) of NdNiO$_2$ in the $(110)$ plane crossing
Nd, Ni, and oxygen vacancy sites as shown in (a).}
\label{fig_nmdft}
\end{center}
\end{figure}

In Figs.~\ref{fig_nmdft}~(b) and (c) we present a typical result
of a non-spin-polarized GGA calculation on NdNiO$_2$,
in a form of atomic orbital projected band dispersion diagrams.
This should give us information about the directional dependence of the dispersion of the various atomic like states,
which is the first step in the evolution of a model Hamiltonian approach to the problem.
As already mentioned, two electron pocket bands cross the Fermi
level beside the cuprate-like $3d_{x^2-y^2}$ band,
one with a strong Nd~$5d_{3z^2-r^2}$ character
at $\Gamma$ and one with mixed Nd~$5d_{xy}$ and
Ni~$3d_{3z^2-r^2}$ characters at $A$.
Most interesting, though, is the fact that the Ni~$3d_{3z^2-r^2}$ states
are found to behave in
a rather non-trivial way\cite{Choi20}. Thus, while along the $Z-R-A-Z$ $k$-path,
which belongs to the $k_z=\pi$ Brillouin zone boundary,
the Ni~$3d_{3z^2-r^2}$ band is flat and lies fully below the Fermi level at around $-1$~eV,
it then strongly disperses along the $c$ direction going from $Z$ to $\Gamma$
where it stops at around $-3$~eV and roughly stays there in the entire
$k_z=0$ plane (the $\Gamma-X-M-\Gamma$ sector).
Moreover, at all the $k$-vectors with $k_z\ne \pi$ the Ni~$3d_{3z^2-r^2}$ states
become
scattered over multiple bands in a wide energy range
both below and above the Fermi level, which signals their partial occupation
and ensuing active involvement in low-energy electronic excitations.

The near flat band behaviour in the variation of the in-plane $k$ for
$k_z=0$ or $k_z=\pi$ suggests that the band has a strong
one-dimensional (1D) character.
Remarkably, it becomes clear from
inspecting the zeronium local orbital projections
in Fig.~\ref{fig_nmdft}~(d) that
such 1D like behaviour is a result of the strong mixing of
the Ni~$3d_{3z^2-r^2}$ states with the interstitial states in the $c$ axis direction.
Indeed, one finds that, on one hand,
the zeronium $s$ orbital has a large contribution
to the occupied band at around $-3$~eV,
which is highlighted by a yellow rectangle in the figure,
at $k$-points with $k_z=0$ (the $\Gamma-X-M-\Gamma$ sector). 
This is also the band that has a strong Ni~$3d_{3z^2-r^2}$ character
in Fig.~\ref{fig_nmdft}~(c).
On the other hand, at the $k$-points with $k_z=\pi$ (the $Z-R-A-Z$ sector),
the Z~$s$ orbital character is mostly above the Fermi level,
while the Ni~$3d_{3z^2-r^2}$ orbital character is fully concentrated in
the occupied band at around $-1$~eV. This is consistent with the fact that,
by symmetry, an $s$ orbital and a $3z^2-r^2$ orbital that alternate along
the $z$ direction have the strongest coupling when $k_z=0$ and
zero coupling when $k_z=\pi$.
Hence, the band at $-3$~eV enclosed in the yellow rectangle corresponds
to the bonding combination of the Ni~$3d_{3z^2-r^2}$
and Z~$s$ orbitals,
whereas their anti-bonding combination is pushed above the Fermi
level where it hybridizes with empty Nd~$5d$ orbitals.
This also explains why in the $Z-R-A-Z$ sector the Ni~$3d_{3z^2-r^2}$
band at $-1$~eV appears flat and well disentangled from other
bands, because this is where $k_z=\pi$ and its coupling
with the Z~$s$ states is canceled out. Moreover, because
of the large radial extend of the Ni~$3d_{3z^2-r^2}$ orbital in the $z$ directions
and small extent in the plane,
the hybridization in the in-plane directions will be relatively small. Note that coupling of Z~$s$
to any of the other four Ni~$3d$ orbitals is symmetry
forbidden at any $k$-vector.
As shown in Supplementary Material\cite{sm},
very similar band structures with the same
zeronium features are obtained for
PrNiO$_2$\cite{Osada_2020} and LaNiO$_2$\cite{Osada_2021},
two other parent IL nickelates
demonstrating superconductivity upon doping.

\begin{figure}
\begin{center}
\colorbox{white}{
\includegraphics[width=0.47\textwidth]{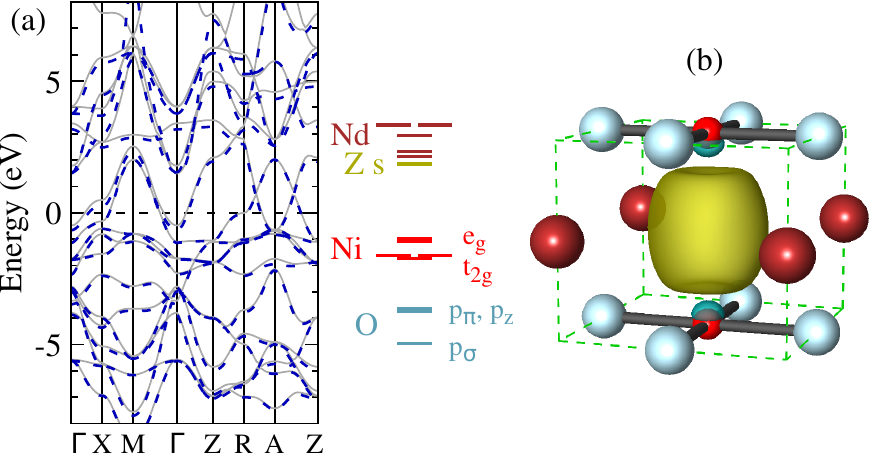}
}
\caption{(a) GGA (solid grey) and TB model (dashed blue)
band dispersions of NdNiO$_2$. On the right-hand side also shown
is the energy-level diagram with the on-site energies
of the 17 atomic and electride orbitals entering our TB model.
The Fermi energy is at zero.
(b) The zeronium~$s$ Wannier function.}
\label{fig_wannier}
\end{center}
\end{figure}

\begin{table}
\caption{
The on-site energies in eV
of the 17 orbitals entering our MLWF based TB model.
The Fermi energy is at zero.
}
\begin{center}
\begin{tabular}{llllll}
\hline\hline
&&Z~$s$&1.84&&\\
\multicolumn{2}{c}{Ni}&&& \multicolumn{2}{c}{Nd}\\
$3d_{xy}$       &-1.74 &&& $5d_{xy}$       &2.15\\
$3d_{xz,yz}$    &-1.61 &&& $5d_{xz,yz}$    &3.34\\
$3d_{x^2-y^2}$  &-1.01 &&& $5d_{x^2-y^2}$  &2.93\\
$3d_{3z^2-r^2}$ &-1.08 &&& $5d_{3z^2-r^2}$ &2.33\\
&&\multicolumn{2}{c}{O}&&\\
&&$2p_{\sigma}$ &-4.96&&\\
&&$2p_{\pi}$    &-3.74&&\\
&&$2p_{z}$      &-3.64&&\\
\hline\hline
\end{tabular}
\end{center}
\label{table_e}
\end{table}

\begin{table}
\caption{
The nearest-neighbor electron hopping integrals
in eV in our MLWF based TB model.
Only those with the absolute values larger than 0.3~eV are listed.
The $\sigma$ and $\pi$ designations of the O~$2p$ orbitals
are done considering their position relative to a neighboring Ni atom.
Note that there is no Nd~$5d$\texttwelveudash Ni~$3d$ hybridization in this list,
which indicates that the dominant interaction path
between these two orbitals is through Z~$s$.}
\label{table_t}
\begin{center}
\begin{tabular}{lrllr}
\hline\hline
Z~$s$\texttwelveudash Ni~$3d_{3z^2-r^2}$ &-1.14 & \hspace{1cm} &Ni~$3d_{x^2-y^2}$\texttwelveudash O~$2p_{\sigma}$  & 1.32\\
Z~$s$\texttwelveudash Nd~$5d_{xy}$ &-1.04 & \hspace{1cm} & Ni~$3d_{3z^2-r^2}$\texttwelveudash Z~$s$ &-1.14\\
Z~$s$\texttwelveudash Nd~$5d_{3z^2-r^2}$ &0.66 & \hspace{1cm} & Ni~$3d_{xz}$\texttwelveudash O~$2p_{z}$            & 0.78\\
Z~$s$\texttwelveudash O~$2p_{\sigma}$ &0.66 & \hspace{1cm} &  Ni~$3d_{xy}$\texttwelveudash O~$2p_{\pi}$          & 0.71\\
Z~$s$\texttwelveudash O~$2p_z$ &0.65 & \hspace{1cm} & Ni~$3d_{3z^2-r^2}$\texttwelveudash O~$2p_{\sigma}$ & 0.67\\
\\
Nd~$5d_{xz,yz}$\texttwelveudash O~$2p_{\pi}$ & 1.22 & \hspace{1cm} & O~$2p_{\sigma}$\texttwelveudash O~$2p_{\sigma}$ & 0.40\\
Nd~$5d_{xy}$\texttwelveudash Z~$s$ & -1.04 & \hspace{1cm} & O~$2p_{\pi}$\texttwelveudash O~$2p_{\pi}$ &-0.36\\
Nd~$5d_{xz,yz}$\texttwelveudash O~$2p_z$ & 0.96 \\
Nd~$5d_{x^2-y^2}$\texttwelveudash O~$2p_z$ & 0.88 \\
Nd~$5d_{xy}$\texttwelveudash O~$2p_{\sigma}$ & 0.56 \\
Nd~$5d_{3z^2-r^2}$\texttwelveudash O~$2p_{\pi}$ & 0.53 \\
Nd~$5d_{xz,yz}$\texttwelveudash O~$2p_{\sigma}$ & 0.50 \\
Nd~$5d_{x^2-y^2}$\texttwelveudash O~$2p_{\pi}$ & 0.41 \\
Nd~$5d_{3z^2-r^2}$\texttwelveudash O~$2p_{z}$ & -0.32 \\
\hline\hline
\end{tabular}
\end{center}
\end{table}

We also perform an MLWF analysis
of the NdNiO$_2$ bandstructure using
the five Nd~$5d$ orbitals, the five Ni~$3d$ orbitals,
the three $2p$ orbitals of each one of the two oxygens in the primitive unit cell,
and the Z~$s$ orbital as a
basis of the corresponding effective tight-binding (TB) model.
We note that several other studies have been reported
that
also consider
an empty $s$-symmetry interstitial orbital
centered at the O vacancy sites\cite{Gu20,Nomura19,Xie22,Ding23}
but here we focus on discussing the important ways it plays a role
in the physical interpretation of the band structure of NdNiO$_2$
and especially of the unusual dispersion of the Ni~$3d_{3z^2-r^2}$ band.
The quality of the agreement between our TB model's
and the GGA energy bands is very good, as one can see
in Fig.~\ref{fig_wannier}~(a).
Figure~\ref{fig_wannier}~(b) shows the Z~$s$
Wannier function in real space,
and Tables~\ref{table_e} and \ref{table_t}
provide the on-site energies and the largest nearest-neighbor
hopping integrals in the model.
These parameter values
can be used in model Hamiltonians by transforming them into a configuration interactions like picture and adding in the on-site
Coulomb and exchange interactions as done in the
recent study by Jiang~{\etal}\cite{Jiang22}.
As one can see in Tables~\ref{table_e} and \ref{table_t},
the on-site energies
of the Ni~$3d_{3z^2-r^2}$ and Z~$s$ orbitals are $-1.08$~eV and $1.84$~eV,
respectively, and the nearest-neighbor electron hopping integral between them is -1.14~eV,
which is almost as large
as the $dp\sigma$ hopping of 1.32~eV between the Ni~$3d_{x^2-y^2}$
and O~$2p_{\sigma}$ orbitals.
Interesting also is that the on-site energies of the Nd~$5d$ orbitals
are all higher than that of the Z~$s$ orbital,
as we also demonstrate in the energy-level diagram of
Fig.~\ref{fig_wannier}~(a).
The hybridization of the Z~$s$ orbital with the Nd~$5d_{xy}$
states at higher energies pushes the bonding part down
and it is this that causes the highly mixed band crossing the Fermi energy and forming the electron pocket at $A$.
%

It is important to discuss
the large nearest-neighbor Z~$s$\texttwelveudash Ni~$3d_{3z^2-r^2}$
hopping in the light of the recent hydrogen experiments\cite{Ding23}
that indicate that superconductivity in the Sr substituted
NdNiO$_2$ is only present if there is a substantial amount of H in the structure.
As other theoretical studies have shown\cite{Si20},
that hydrogen assumes a close to H$^-$ ionic state and
favours to be in the O vacancy position.
If this is correct, then we note that the presence of H$^-$ would
block the Z~$s$\texttwelveudash Ni~$3d_{3z^2-r^2}$ hybridization
since the vacancy position is already occupied by a negative charge.
This would put the Ni electronic structure close to that of Cu
in the cuprates. One could also speculate that the inclusion of some
F$^-$ in the vacancy position could do the same and perhaps more efficiently.
This is worth a try but one should also realize that the
relatively large size of F$^-$ could cause a substantial $c$ axis
expansion which may indeed act to destabilize the structure.


\begin{figure}
\begin{center}
\colorbox{white}{
\includegraphics[width=0.47\textwidth]{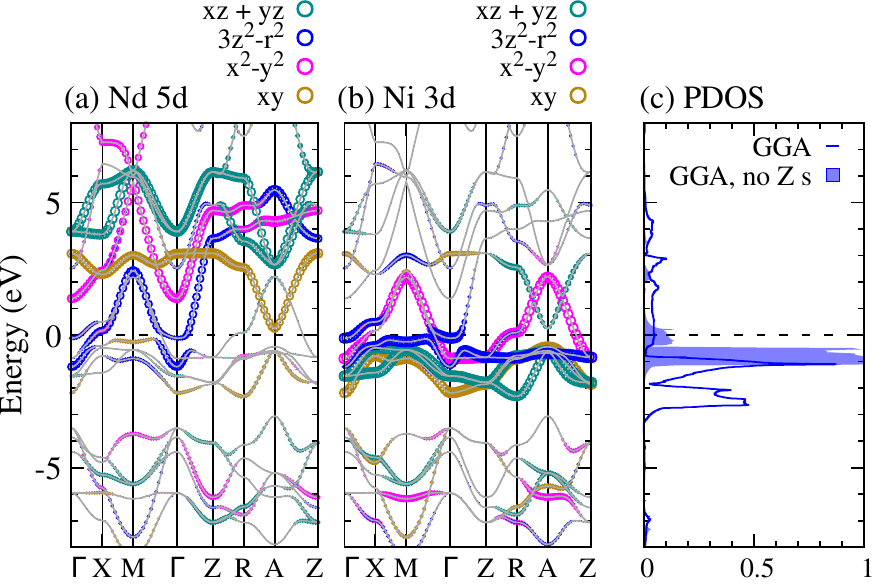}
}
\caption{The result of a computational experiment where the zeronium~$s$
orbital is forcefully depopulated yielding significant changes
in the NdNiO$_2$ electronic structure around the Fermi level:
(a) and (b) Atomic orbital projections of the altered band structure.
(c) A comparison between the Ni~$3d_{3z^2-r^2}$ projected densities of states
(PDOS) of NdNiO$_2$ before (``GGA``) and after (``GGA, no Z~$s$``)
depopulating the Z~$s$ orbital.
The Fermi energy is at zero.}
\label{fig_exp}
\end{center}
\end{figure}

\subsection{Computational experiment}
In contrast with the emphasis
on the Nd~$5d$ states that prevails in the literature,
our results from Table~\ref{table_t}
indicate that their direct hybridization
with the Ni~$3d_{3z^2-r^2}$ and Ni~$3d_{x^2-y^2}$
orbitals is small, but it is rather
the Z~$s$ orbital that mediates the involvement of these
empty states with the Ni states.

To better illustrate this important observation,
we consider here results from
a computational experiment where the Z~$s$ orbital is
intentionally removed
from playing any role in the low-energy electronic properties.
This can be achieved via a charge self-consistent calculation
where electrons are made to pay a high energy cost for occupying this orbital.
The resulting bandstructure shown in Figs.~\ref{fig_exp}~(a) and (b)
displays significant differences with the original one in Figs.~\ref{fig_nmdft}~(b)
and (c).
Most notably, the Ni~$3d_{3z^2-r^2}$ band becomes essentially flat
throughout the entire Brillouin zone.
Its dramatic transformation is emphasized in Fig.~\ref{fig_exp}~(c)
showing the Ni~$3d_{3z^2-r^2}$ projected densities of states before and after
the removal of the Z~$s$ orbital. Furthermore,
instead of two electron pocket bands, we now have only one at $\Gamma$,
of mainly Nd~$5d_{3z^2-r^2}$ character.
It crosses and hybridizes with the Ni~$3d_{3z^2-r^2}$ band,
but this hybridization is much weaker compared
to that between the Ni~$3d_{3z^2-r^2}$ and Z~$s$ bands,
in accordance with Table~\ref{table_t}.
In essence, the Ni~$3d_{3z^2-r^2}$ orbital becomes much more atomic-like here, while it
clearly cannot be treated as such in reality because of its strong
covalent bonding with the zeronium~$s$ orbital.

\subsection{ELF (electron localization function)}
We find a total of 0.36 electride electrons inside a sphere of radius 0.85~{\AA}
around the vacancy site, out of which 0.24
occupy the local $s$ orbital. This is large enough to support
the electride scenario, but (as it is often the case in studies
on electrides \cite{Dale18}) not sufficient to produce a significantly
visible maximum in the real-space electron density distribution,
due to the strong background density of other electrons.
There is, however, a better way of measuring electron localization
in an electride material which is by using the electron localization
function (ELF) \cite{Becke90,Savin92}. ELF has been established as
one of the best descriptors for identifying electride systems \cite{Dale18}.
The definition of ELF
is tailored such that its value can vary between 0 and 1, with 1 corresponding
to maximal electron localization.
As one can see in Fig.~\ref{fig_nmdft}~(e), the computed ELF of NdNiO$_2$
has a clear maximum
at the oxygen vacancy ({\ie}, zeronium) site. A maximum of a similar strength was
taken as a sign of an electride-like nature of the prototypical
electride compound  [Ca$_{24}$Al$_{28}$O$_{64}$]$^{4+}$($4e^-$) \cite{Matsuishi03}.

There is also a clear minimum in the ELF of NdNiO$_2$
at a distance of about 1~{\AA}
going from the Ni site towards the Z site.
This further indicates that
the electron density in that region
cannot be derived solely from the Ni~$3d$
wave function but rather from a combination
of the Ni~$3d$ and Z~$s$ wave functions, because
a pure atomic $3d$ wave function does not have a radial node.
An alternative interpretation would be to say
that this minimum is a result of
an effective mixing of the Ni~$3d$ states with the $4d$ states,
a point of view adopted, for instance, by Katukuri {\etal} in Ref.~\onlinecite{Katukuri20}.
The $4d$ wave function of an isolated Ni$^+$ ion
has its node indeed close to 1~{\AA} from the Ni nucleus.
This observation is of great importance in the development of model Hamiltonians to describe these correlated electron systems and,
as we will argue below,
might also help us understand better the results of DFT+U calculations.

\begin{figure}
\begin{center}
\colorbox{white}{
\includegraphics[width=0.47\textwidth]{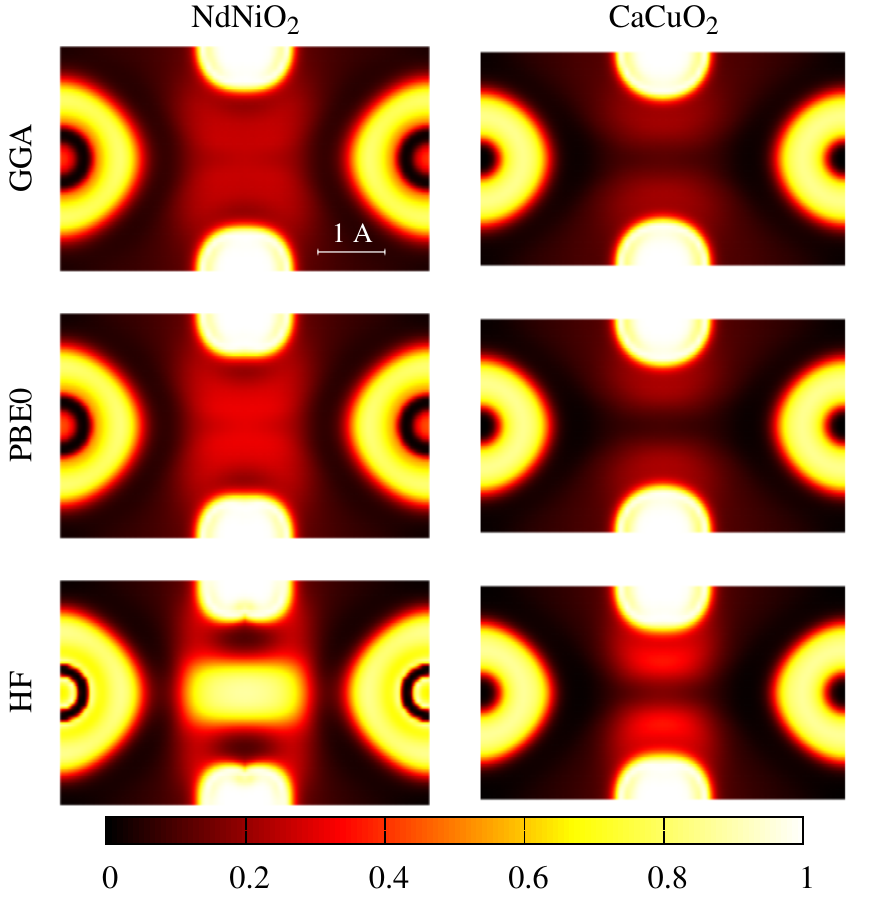}
}
\caption{Electron localization functions of NdNiO$_2$ and CaCuO$_2$
in the (110) crystal plane,
calculated using the GGA, PBE0,
and Hartree-Fock (HF) methods
and the pseudo-potential code Quantum Espresso.}
\label{fig_elf}
\end{center}
\end{figure}

It is interesting to compare the ELF of NdNiO$_2$
with that of the iso-structural cuprate system CaCuO$_2$.
The two top panels of Fig.~\ref{fig_elf}
show their ELFs obtained from ferromagnetically
spin-polarized GGA calculations using the pseudo-potential code
Quantum Espresso. We will explain the need to switch to
both a new code and a new magnetic configuration
below, but for now let us note that this
does not change the ELFs of the considered oxides
in the region around the zeronium, as one can see
by comparing with Fig.~\ref{fig_nmdft}~(e).
An important result is that, opposite to NdNiO$_2$,
the ELF of CaCuO$_2$ has a minimum rather than a maximum at the oxygen vacancy site.
This indicates that the electride-like electronic behavior
of the IL nickelates is a unique feature
that makes them distinctly different from
the cuprates.

Moreover, this difference becomes more and more pronounced
as one switches to using a hybrid functional with some
amount of an exact [{\ie}, Hartee-Fock (HF)] exchange in it,
a strategy known to help improve the description of electronic correlations
and electron localization,
particularly, in the case of electride electrons
(see Ref.~\onlinecite{Dale18}, page 9373).
We consider the PBE0 hybrid functional method \cite{pbe0} which is based on
the following simple expression for the exchange
(''x``) and correlation (''c``) energy functional:
\[
E^{\text{PBE0}}_{\text{xc}} = \frac{1}{4} E^{\text{HF}}_{\text{x}}
+
\frac{3}{4} E^{\text{GGA}}_{\text{x}}
+
E^{\text{GGA}}_{\text{c}},
\]
in which a quarter of the exchange term of the original GGA functional
is proportionally replaced by an HF contribution.
As one can see in the two middle panels of Fig.~\ref{fig_elf},
introducing a quarter of an exact exchange results
in an enhanced localization of the electride electron in
NdNiO$_2$ but does little in the case of CaCuO$_2$,
which reinforces our previously stated conclusions.
Finally,
performing a pure HF calculation (the bottom panels of Fig.~\ref{fig_elf})
is found to drive the electride electron of NdNiO$_2$ towards full
localization.

Before finishing this section, let us explain our choices
regarding the magnetic configuration and the DFT code
used to produce the results in Fig.~\ref{fig_elf}.
First, it becomes very challenging to stabilize a non-magnetic solution
as soon as one adds some amount of an exact exchange into the energy functional,
and therefore all the calculations presented in Fig.~\ref{fig_elf}
are performed keeping consistently the same ferromagnetically ordered Ni spin configuration,
which is the only possible configuration one can have in a primitive unit cell.
Second,
in the case of hybrid functional calculations,
computational efficiency becomes a serious issue
which can be better dealt with in a
pseudo-potential code like Quantum Espresso.
For the same reason, we did not choose to consider here some kind
of an antiferromagnetic order, as this would require using
a computationally more demanding multiple formula unit supercell.

\subsection{Can one use the DFT+U method to unveil the
Fermi surface of the IL nickelates?}
\begin{figure*}
\begin{center}
\colorbox{white}{
\includegraphics[width=0.94\textwidth]{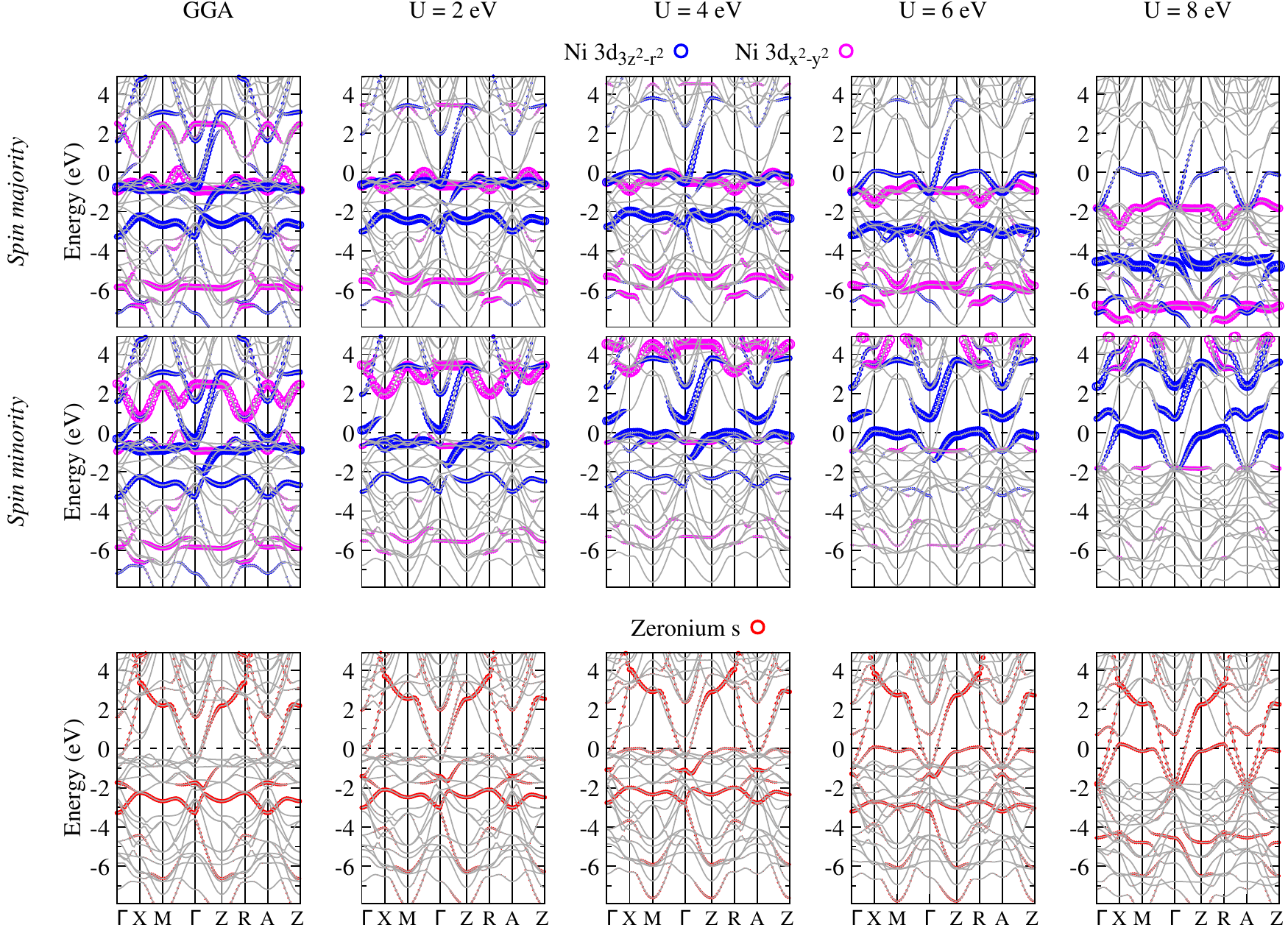}
}
\caption{The band structure of NdNiO$_2$
in the $G$-type antiferromagnetically ordered state
obtained from WIEN2k with Ni atomic
and electride orbital projections as a function
of $U$ in the DFT+U method. The Fermi energy is at zero.}
\label{fig_bandsu}
\end{center}
\end{figure*}

We now explore further how magnetism and electronic correlations
might affect
the behavior of the electride electrons in the parent IL nickelates.
Our hybrid functional calculations
already indicate that
correlations can strongly enhance their electride-like properties,
but a detailed study would require us to use a more computationally flexible method.
Therefore, we will consider here the popular
DFT+U approximation \cite{Anisimov91},
in which local electronic correlation effects are
treated in a static mean-field fashion.
This is a method that has extensively been used in computational studies of
various TM compounds
with strong on-site electronic correlations,
including the IL nickelates\cite{Liu_2020,Been21,Hepting20}.
Its efficiency, however,
comes at a price. First, with DFT+U one is bound to
deal with a magnetically long-range ordered state,
which is different from the experimentally observed ground state of the parent
IL nickelates where localized magnetic moments, although
indeed present, have no long-range correlations\cite{Fowlie22}.
Second, in DFT+U the electron interaction parameters
appear as adjustable parameters, introducing additional uncertainty. There are also
other, more subtle yet potentially very important, differences in how a DFT+U calculation
can be set up, which will be discussed later.

Figure~\ref{fig_bandsu} shows the evolution
of the DFT+U band structure of NdNiO$_2$
as a function of $U$, the on-site Coulomb repulsion between Ni~$3d$
electrons.
The Hund’s rule exchange coupling, $J_{\text{H}}$,
is set to zero for simplicity.
We note, however, that this does not mean
that the net $J_{\text{H}}$ is zero, because
in DFT full exchange is included already at the GGA level
but with a Hamiltonian involving only $S_z\cdot S_z$
terms. In other words, the Hund's rule exchange is treated as Ising-like
since DFT methods with $\vec{k}$ as a good quantum number have only $S_z$ and not the local $S$ as a good quantum number. This is very important since the singlet energy in
$J\vec{S}\cdot\vec{S}$ is $-3/4J$ for spin 1/2 and the triplet energy
is $+1/4J$, while for the Ising-like interaction the spin parallel and antiparallel
energies are $\mp1/4J$.

We order the Ni spin magnetic
moments antiferromagnetically with a $Q=(\pi,\pi,\pi)$ ({\ie}, $G$-type)
ordering vector, which causes doubling of the unit cell
and, consequently, of the number of bands compared with Figs.~\ref{fig_nmdft}~(b)-(d).
As mentioned earlier,
the need to have a long-range magnetic order is
an unavoidable component of DFT+U,
reflecting the static nature of this approach.
However, as for example discussed by Trimarchi and co-workers\cite{Trimarchi18},
the influence of the kind of magnetic order usually is very small especially if there are resulting strongly localized spins in the ordered state.

Several comments are to be made regarding
the antiferromagnetic GGA results presented in the left-most column of Fig.~\ref{fig_bandsu}.
First, like in the cuprates,
spin polarization occurs involving mainly the Ni~$3d_{x^2-y^2}$
bands.
Second, this has almost no effect on the Z~$s$
bands, including the occupied ones at -3~eV,
which means that the electride-like nature of the compound stays intact.
The small electron pockets at $\Gamma$ and $A$ (now folded on top of each other)
are also preserved. But now, because of the antiferromagnetic
splitting of the Ni~$3d_{x^2-y^2}$
bands, the compensating positive charge is located in the similarly small
hole pockets at half way between $M$ and $\Gamma$ (and, equivalently, $A$ and $Z$) and has
a spin majority Ni~$3d_{x^2-y^2}$ orbital character.
This suggests that, within the rigid band shift approximation,
in the chemically hole-doped superconducting nickelates like Sr$_{x}$Nd$_{1-x}$NiO$_2$,
doped holes should bind with the stoichiometric holes
into local singlets with the $x^2-y^2$ symmetry,
a situation similar to what happens in the cuprates\cite{Jiang20,Jiang22}.
Note that the occupied $x^2-y^2$ states are strongly
hybridized with the same symmetry molecular combination of
O~$2p$ orbitals as in the Zhang-Rice picture,
which is obvious from the $x^2-y^2$ symmetry states forming another band
at between 5 and 6~eV of mainly O~$2p$ molecular orbital character.

This picture, however, changes into a qualitatively very different one
as the value of $U$ starts to increase. First, the Ni~$3d_{3z^2-r^2}$ electrons
become spin-polarized too and get coupled ferromagnetically
with the Ni~$3d_{x^2-y^2}$ electrons. Second,
only one electron pocket survives. Also,
instead of the spin {\it majority} Ni~$3d_{x^2-y^2}$ hole pocket between $M$
and $\Gamma$,
we have now for $U>4$~eV a shallow hole pocket at $X$ (and, equivalently, $R$)
with a spin {\it minority} Ni~$3d_{3z^2-r^2}$ orbital character.
In this situation, the chemically doped holes and the stoichiometric holes
would be found forming local triplets rather than singlets,
which could possibly lead to
a superconducting state with a completely different symmetry and
perhaps even of a fundamentally different nature.
Interestingly, the Z~$s$ orbital will play
an important role in this scenario as at large $U$ it acquires
a substantial weight in the hole pocket band at $X$
resulting from its strong hybridization with the Ni~$3d_{3z^2-r^2}$ orbital.

\begin{figure}
\begin{center}
\colorbox{white}{
\includegraphics[width=0.45\textwidth]{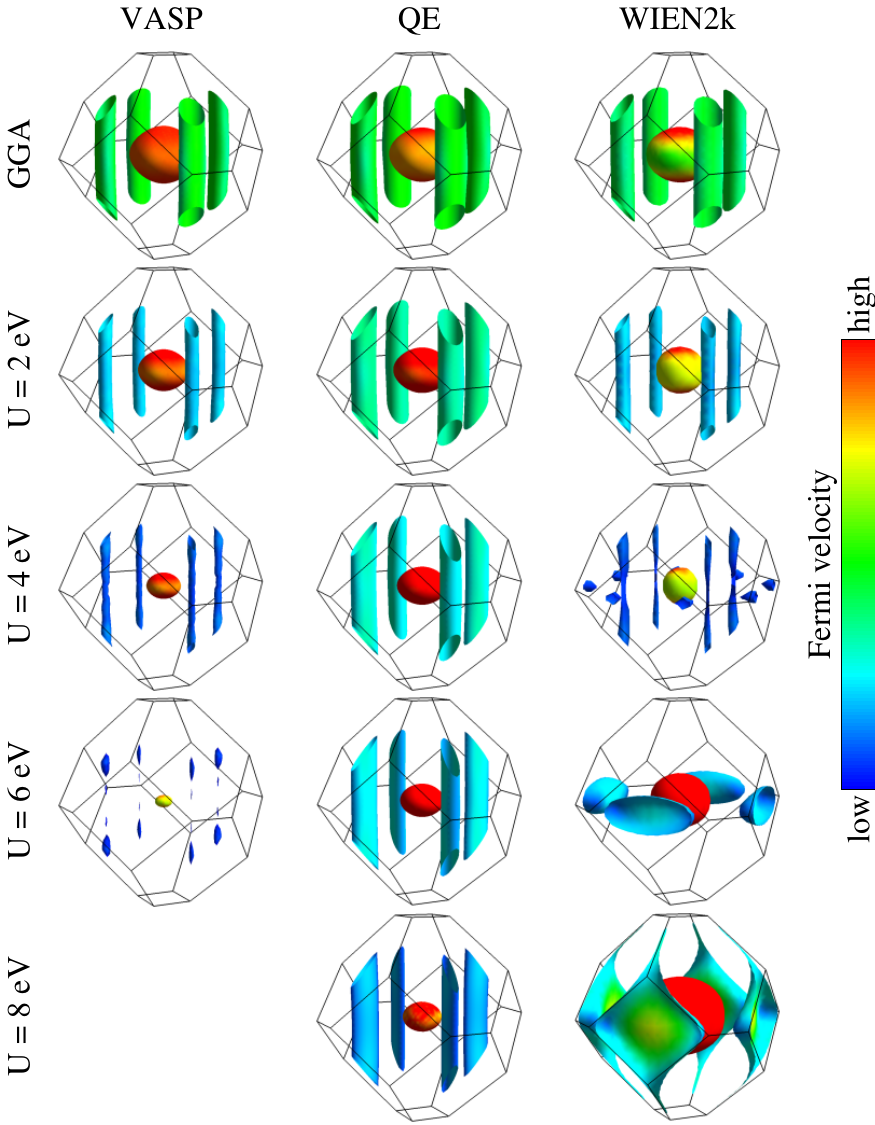}
}
\caption{Fermi surface of NdNiO$_2$ calculated in DFT+U
using different DFT codes for a range of on-site
Hubbard interaction parameters $U =$ 0, 2, 4, 6, and 8~eV.
The Hund's exchange coupling is set to zero. The Fermi surface
color represents the Fermi velocity.}
\label{fig_fs}
\end{center}
\end{figure}

We have noticed
that some of the published DFT+U based studies\cite{Liu_2020,Been21,Hepting20}
of the IL nickelates
report a qualitatively different $U$-dependence
of the electronic structure, finding in particular
that there is a charge gap opening
at $U=8$~eV, whereas in our DFT+U calculations NdNiO$_2$
remains metallic at all $U$.
Suspecting
that the difference in the choice of a DFT code
might be the issue,
we repeat our calculations using also the pseudo-potential codes VASP
and Quantum Espresso.
The obtained results as a function of $U$ and a DFT code
are summarized in Fig.~\ref{fig_fs} in a form of Fermi surfaces.
Indeed, the three considered DFT codes are found to give qualitatively different $U$-dependence
for the Fermi surface of NdNiO$_2$ and, as such, for its electronic structure overall.
%

We stress that within each code we have intentionally
chosen to use {\it nominally}
the same rotationally invariant
formulation of the DFT+U method by Liechtenstein {\it et al.} \cite{Liechtenstein95}
(which is often referred to as a ''self-interaction correction`` method).
Therefore, the main source for the discrepancy
has to be the different ways that the codes
use to define local Ni~$3d$ orbitals for
calculating occupation matrices in DFT+U expressions.
By default, atomic and Lowdin orthogonalized atomic wave
functions are used in QE and VASP, respectively,
whereas WIEN2k considers atomic-like wave functions that go to zero
outside muffin tin spheres.
These differences have minor effect when the correlated
orbitals in question are well-localized (see, for instance,
corresponding results for CaCuO$_2$ in Supplementary Material\cite{sm}),
but become critically important in the presence of strong
hybridization with uncorrelated subset.
In the case of the IL nickelates, such is
primarily the hybridization between
the Ni~$3d_{3z^2-r^2}$ and Z~$s$ orbitals.
One way to see this
is by noticing that it is the hole pocket
band of exactly that mixed Ni~$3d_{3z^2-r^2}$ and Z~$s$ orbital character
that gets pinned to the Fermi level
at $U>4$~eV in WIEN2k and whose absence results in
the charge gap opening in VASP.
This strong hybridization is all but absent in the cuprates,
while both have relatively strong hybridization of
the O~$2p$ and the TM $3d_{x^2-y^2}$ orbitals although weaker for the nickelates because of the larger charge-transfer energy.

It is also worth
mentioning that in WIEN2k projections onto local states of a certain symmetry
are done without considering the radial dependence, {\ie},
without differentiating between principal quantum numbers,
while in VASP and Quantum Espresso this is not the case,
which, in the light of our earlier argument
about the possible mixing of Ni~$3d$ and $4d$ states,
can be another important factor contributing to the discrepancy.

Summarizing this section, Fig.~\ref{fig_fs} provides yet another demonstration
of the important role
of the Z~$s$ orbital
coupled with the Ni~$3d_{3z^2-r^2}$
orbital in a strong covalent bond.
Such unique property of the IL nickelates combined with the discussed
limitations of the DFT+U method
causes an uncertainty in the nature of the
symmetry of the hole pocket and the states that are closest to the Fermi level.


\section{Discussion}
\subsection{Ni electronic configuration: $d^9$, $d^8s$, or $d^9s\underline{L}$?}
Our presented DFT based analysis of the parent IL  nickelates points to a microscopic
picture where
the oxygen vacancy centered $s$ orbital
hybridizes very strongly with the Ni~$3d_{3z^2-r^2}$ orbital, pulling a large portion
of its atomic electron density into the interstitial region.
As a result, these two orbitals
form a bonding and an anti-boning states,
with the bonding state having larger Ni~$3d$ than Z~$s$
character and being totally occupied
and with the anti-bonding state having more Z~$s$ character
and being totally empty.
This leads to a substantial hole concentration in the
Ni~$3d_{3z^2-r^2}$ band driving Ni$^+$ towards Ni$^{2+}$.
Noteworthily, a series of recent DFT+DMFT studies
have arrived at a similar conclusion about the Ni ionisation state
\cite{Lechermann20, Lechermann20a,Leonov20,Xie22,
Petocchi20, Wang20, Pickett21, Werner20,Ortiz21,Karp21}.
Such deviation from
the nominal valency is reminiscent of that encountered in
negative charge-transfer compounds \cite{Zaanen85,Sawatzky_Green_book},
but here the role of ligand states pulling in holes from the transition metal states is delegated to the zeronium s states pulling in electrons.
Many materials
with rich physics belong to this class, including the high-$T_{\text {c}}$
superconductors bismuth oxides \cite{Sleight75,Foyevtsova15},
the battery material LiNiO$_2$ with a proposed electronic
high-entropy state \cite{Goodenough58,Foyevtsova19}, and also the previously
mentioned precursor compounds $R$NiO$_3$ \cite{Medarde97}. In $R$NiO$_3$,
for instance,
a negative value of the charge-transfer energy between
the Ni~$3d$ and O~$2p$ states
is known to result in a $d^8\underline{L}$ (Ni$^{2+}$) rather than the formal
$d^7$ (Ni$^{3+}$)
configuration for Ni ions, where $\underline{L}$ stands for an O~$2p$ hole coupled
strongly in a covalent bond with a Ni~$3d$ hole.
By analogy, we can say that in the IL nickelates
a state like $d^8s$, with $s$ denoting an
electron in the Z~$s$ orbital bound in a strong covalent bond with
neighboring Ni~$3d_{3z^2-r^2}$ orbitals, should have a substantial
weight in the ground state Ni configuration.

It is important to keep in mind
that in order to fully reveal the true ground state configuration of Ni,
including the local spin and multiplet structure, one needs to go
beyond DFT and use advanced many-body techniques.
In this regard, recent
configuration interaction based impurity calculations
by Jiang~{\etal}\cite{Jiang22}
show that the ground state is a mixture of
similarly strong contributions from
the $d^9_{x^2-y^2}$, $d^{8,\, S=1}_{x^2-y^2,\, z^2}\, s$,
$d^9_{z^2}\, \underline{L}_{x^2-y^2}\, s$,
and $d^{10}\underline{L}_{x^2-y^2}$ configurations,
all having a net spin of $1/2$.
This is very different from the cuprates
where the
ground state electronic structure is dominated by
$d^9_{x^2-y^2}$ and $d^{10}\underline{L}_{x^2-y^2}$.

Our findings complement the ideas from Ref.~\onlinecite{Ding23} about the
possibility of the interstitial charge suppressing superconductivity
in hydrogen-poor samples. We emphasize, however,
that it is the nearest-neighbor Z~$s$\texttwelveudash Ni~$3d_{3z^2-r^2}$
hybridization that is primarily responsible
for turning the stoichiometric nickelates into effectively
three-dimensional systems and thus driving them
away from the two-dimensional regime that is known to favor
superconductivity in cuprates. This is supported by
the calculations for the 20 percent hole doped LaNiO$_2$ system
presented in Supplementary Material\cite{sm},
in which the electride-like nature
and the associated strong out-of-plane electronic coupling
are found to be well preserved.

Important implications regarding magnetism
are also in place. The on-site Hund's rule coupling dictates
that the Ni~$3d_{3z^2-r^2}$ holes have the same spin as the $d_{x^2-y^2}$ holes,
again pointing to a strong admixture of the $d^8$ ($S=1$) state in the net
ground state configuration of Ni.
This is indeed what our DFT+U calculations in Fig.~\ref{fig_bandsu} suggest
and what the mentioned DFT+DMFT studies report
and is in line with the conclusions of the impurity calculations including all multiplets by Jiang~{\etal}\cite{Jiang22}
We can also think of this as a large
part of the Ni$^+$~$3d_{3z^2-r^2}$ electron finding itself in the zeronium
site with spin parallel to the local $d_{x^2-y^2}$ remaining electron.
This would result in a very strong anti-ferromagnetic coupling
between the Ni spins in the one-dimensional chains along the $c$ direction,
whose consequences require a detailed theoretical exploration,
especially, in connection with the low-energy
excitations seen in the recent resonant
inelastic x-ray scattering (RIXS) experiments
\cite{Rossi21_xas,Lu21}, but
this kind of a study is beyond
the scope of the present work.

\subsection{Possible excitonic insulating
state at low temperatures}
Keeping in mind that, strictly speaking, the standard DFT+U approach
is not applicable to the IL nickelates,
let us nevertheless note
that all the three DFT codes
are consistent in that
they find a strongly dispersive electron pocket
band and a much flatter hole pocket band
at physically reasonable $U$ values around 6~eV.
Furthermore, the number of charge carriers occupying these
electron and hole pockets is extremely small.
In the specific case
of the WIEN2k solution with $U=6$~eV (Fig.~\ref{fig_bandsu}),
we find that this number corresponds
to a charge carrier concentration of only $0.031$~electrons and holes per formula unit.
This is a situation where we have a diluted
nearly-free-electron-like gas of electrons
interacting with an equally diluted
pool of holes that are however much more localized and residing primarily on Ni~$3d$ orbitals.
At low enough temperatures, such a system will have a strong tendency
to form  spatially bound excitonic pairs,
which, since the number of electrons and holes has to be equal because of the charge neutrality condition, can
possibly result in an excitonic insulating ground state
for the parent IL nickelates of the kind similar
to that observed in some transition metal dichalcogenides\cite{Wilson77}.
This scenario, which
is in line with the Kondo-like physics
in the limit of the Nozieres exhaustion principle
considered
in Refs.~\onlinecite{Yang22} and \onlinecite{Shao22}, would naturally explain
the measured resistivity up-turn.
Also, in light of this, it is not
impossible that the low-energy
dispersive excitations seen in the recent RIXS
measurements \cite{Rossi21,Lu21}
are of an excitonic origin
rather than being driven by short-range magnetic order excitations.

\section{Conclusion}

In conclusion, the present study elucidates the electride-like nature
of the IL nickelates and reestablishes that in these materials
``Ni$^+$ is not Cu$^{2+}$'' but rather
in a state closer to Ni$^{2+}$ with
substantial $d^{8,\, S=1}_{x^2-y^2,\, z^2}\, s$
and $d^9_{z^2}\, \underline{L}_{x^2-y^2}\, s$
contributions, in addition to the cuprate-like
$d^9_{x^2-y^2}$ and $d^{10}\, \underline{L}_{x^2-y^2}$ ones.
This results from the
strong attractive potential for electrons left behind by the missing O in the NdO layer leading to a considerable charge density attracted to that region in the ground state of the undoped material.
We label this region of the attractive potential
with a fictitious atom with zero nuclear charge (Z)
and find in DFT that its lowest-energy orbital is an $s$ symmetry orbital.

In the language of tight-binding modeling,
such peculiar charge density redistribution
can be described in terms of
the strong hybridization of the Ni~$3d_{3z^2-r^2}$ orbital
with the Z~$s$ orbital
and their resulting bonding/anti-bonding like splitting
that pulls some of the Ni~$3d_{3z^2-r^2}$ electron density into the vacancy region in the ground state.
Our TB analysis indicates also that
there is a strong involvement of the
Nd~$5d_{xy}$
and O~$2p_z$ orbitals, which, however,
have only an indirect effect on the Ni~$3d$ electronic structure.
We also pointed out
that all the theoretical methods considered so far, like DFT, the impurity calculations, and DMFT, lead to the conclusion that the $d^8$
states involved in the ground state are in a triplet configuration and the total configuration has spin 1/2 per formula unit.
However, the propagation of this spin 1/2 composite object is much more complicated than that presented by the corresponding spin 1/2 composite object in the cuprates, which only involves the $x^2-y^2$ holes of
Cu~$3d$ and a linear combination of O~$2p$ orbitals of the same symmetry.
Such distinct difference in the electronic behaviors
of the nickelates and the cuprates is
further emphasized in our calculations of their electron
localization functions, which is a well documented measure of
the electride characteristics.

Our study strongly suggests
that the Z~$s$\texttwelveudash Ni~$3d_{3z^2-r^2}$
covalent bonding
is a necessary component to be explicitly considered in both electronic
structure and model calculations that might eventually help us
unravel the nickelates' elusive Fermi surface
as well as understand the striking sensitivity of their
superconducting properties to hydrogen content.
In the latter regard, we argued
that if the O vacancy orbital is compensated by H$^-$
as recently suggested\cite{Ding23}, this would tend to remove the need for
the Ni~$3d_{3z^2-r^2}$ hybridization with the Z~$s$ state to develop a negative charge in the O vacancy position and so the system would tend to be much more similar to the cuprate case. This suggests that perhaps the resulting superconductivity by the substitution of Nd with divalent Sr and with the substantial
vacancy compensation by H$^-$ would indeed be very similar to that of the cuprates driven mainly by the Ni~$3d_{x^2-y^2}$ orbitals.

Going back to the undoped parent compound though,
we also argued that because of the condition of charge neutrality of the unit cell and the material as a whole in the ground state the much discussed electron pocket is perfectly compensated by a hole pocket of equal size but occurring
in the bands of mainly Ni~$3d_{x^2-y^2}$
or $3d_{3z^2-r^2}$ character. This, in addition
to the missing $d$ electron charge in the Ni$^+$
formal configuration, means that we must have a finite density of
Ni~$d^8$ states, be it with two holes in the $x^2-y^2$ orbital,
which would have to form a singlet as in the Zhang-Rice scenario,
or with one hole in the $x^2-y^2$ and one hole in the $3z^2-r^2$ orbital, which locally would form
an energetically favourable triplet state but for charge neutrality
reasons there must also be another electron around,
{\ie}, Z~$s$, which we find has spin opposite to the $d^8$ triplet
so the net spin is again 1/2.

Concerning the relative likelihoods of the singlet
versus triplet scenarios,
a Lifschitz like transition
from a Fermi surface hole pocket of the $x^2-y^2$ symmetry
into that of the $3z^2-r^2$ symmetry
is found to occur
as a function of $U$ for $U$ values larger than 4~eV
in the DFT+U method and using Wien2k.
This is, however, where we also observed significant
qualitative inconsistencies between different DFT codes, such as Wien2k, VASP,
and Quantum Espresso, which can be traced down
to the complications that the DFT+U method faces
in the presence of a strong hybridization
of correlated orbitals with their uncorrelated surroundings,
which in the present case is exactly the
hybridization between the interstitial Z~$s$
and Ni~$3d_{3z^2-r^2}$ orbitals.

We also pointed out that the increased charge density
at the oxygen vacancy position relative to the usually found
rather uniform electron density in the interstitial region
can be described in a DFT calculation
in several different ways depending
on the basis set of wave functions used.
In the end, as density functional theory dictates
it is only the electron density that matters.
So, the extra density in the Z region
can also be obtained by introducing for example $4d$ orbitals
in the Ni atomic basis,
which have a radial node and have their maximum amplitude
in the vicinity of the O vacancy.
It is the choice of the basis set and the nature of the projections in determining the state that one wants to apply $U$ to that causes
the difference in the Fermi surface for the various codes when applying $U$.
Although beyond the scope of this work, these observations hopefully will
stimulate a more careful look at what causes this strong code dependent Fermi surface in the DFT+U methods.

From the experimental point of view, it would be of great interest
to have a detailed study of the magnetic field and temperature
dependent magnetic susceptibility done
using either x-ray magnetic circular dichroism spectroscopy
or neutron scattering in order to gain insight into the magnetic properties of the IL nickelates.
With regard to the Fermi surface, an
angle-resolved photoemission spectroscopy study
on high-quality stoichiometric crystals would be essential and could in principle
at least distinguish between the $x^2-y^2$ and $3z^2-r^2$ character at
the Fermi level in the doped materials.

\section*{Acknowledgments}
This research was undertaken thanks, in part, to fund-
ing from the Max Planck-UBC-UTokyo Center for Quantum
Materials and the Canada First Research Excellence Fund,
Quantum Materials and Future Technologies Program as well
as from CIFAR and NSERC.

\bibliographystyle{apsrev}

\end{document}